\RequirePackage{ifpdf}

\RequirePackage{fix-cm}

\documentclass[%
floatfix,
showkeys,
nofootinbib, %
superscriptaddress, %
]{revtex4-1}

\ifpdf
  \usepackage{cmap}
\fi
\usepackage{textcomp}

\usepackage[T1,T2A]{fontenc}

\usepackage[utf8]{inputenx}
\input{ix-utf8enc.dfu}

\usepackage[german,russian,english]{babel}

\usepackage{amsmath}
\usepackage{amssymb}

\usepackage{mathtools}
\mathtoolsset{
showonlyrefs,
mathic = true
}

\allowdisplaybreaks

\usepackage{hyperref}
\hypersetup{backref,
 colorlinks=false}
\hypersetup{pdfborder=0 0 0}

\usepackage{microtype}
\UseMicrotypeSet[protrusion]{alltext}

\usepackage{graphicx}

\usepackage[scanall]{psfrag}

\usepackage{listings}
\usepackage{listingsutf8}
\usepackage[frozencache]{minted}

\usepackage{fixltx2e}

\usepackage{nicefrac}

\makeatletter
\def\ps@pprintTitle{%
     \let\@oddhead\@empty
     \let\@evenhead\@empty
     \let\@oddfoot\@empty
     \let\@evenfoot\@oddfoot}
\makeatother

\begin{document}

\graphicspath{{image/modelica-vs-julia/en/}{image/modelica-vs-julia/}{image/}}

\title{Languages for modeling the RED active queue management algorithms: Modelica vs. Julia}

\author{Anna Maria Yu. Apreutesey}
\email{1032193049@pfur.ru}
\affiliation{Department of Applied Probability and Informatics,\\
  Peoples' Friendship University of Russia (RUDN University),\\
  6 Miklukho-Maklaya St, Moscow, 117198, Russian Federation}

\author{Anna V. Korolkova}
\email{korolkova-av@rudn.ru}
\affiliation{Department of Applied Probability and Informatics,\\
  Peoples' Friendship University of Russia (RUDN University),\\
  6 Miklukho-Maklaya St, Moscow, 117198, Russian Federation}

\author{Dmitry S. Kulyabov}
\email{kulyabov-ds@rudn.ru}
\affiliation{Department of Applied Probability and Informatics,\\
  Peoples' Friendship University of Russia (RUDN University),\\
  6 Miklukho-Maklaya St, Moscow, 117198, Russian Federation}
\affiliation{Laboratory of Information Technologies\\
  Joint Institute for Nuclear Research\\
  6 Joliot-Curie, Dubna, Moscow region, 141980, Russian Federation}

\begin{abstract}
This work is devoted to the study of the capabilities of the Modelica and Julia programming languages for the implementation of a continuously discrete paradigm in modeling hybrid systems that contain both continuous and discrete aspects of behavior. A system consisting of an incoming stream that is processed according to the Transmission Control Protocol (TCP) and a router that processes traffic using the Random Early Detection (RED) algorithm acts as a simulated threshold system.

\end{abstract}


\maketitle

\section{Introduction}
\label{sec:intro}

The possible methods for investigating complex systems are the
construction of a discrete event model, the construction of a
continuous model, and hybrid
modeling~\cite{kulyabov:2016:ecms:red}. Such hybrid systems combine
the operation of continuous and discrete elements, for example,
systems with a discrete control device and a control object with a
continuous nature of operation~\cite{farnqvist:2002:hybrid_modeling}. 
The system for the further research is the
hybrid model of the interaction of the data transmission process via
the Transmission Control Protocol (TCP) and the Random Early Detection
(RED) algorithm as an algorithm of regulating the flow state. When
modeling TCP-like traffic, the liquid (continuous) approach can be
used, but it is especially important to consider discrete transitions
between TCP states and a discrete packet drop function in RED-type
algorithms. It is the hybrid approach that reflects these important
features of the simulated system.

  \subsection{The structure of the article }
\label{sec:structure}

The structure of the article is as following. 
The~\ref{sec:RED}~section describes the basic principles 
and the mathematical model of the operation of the congestion control 
mechanism in the TCP protocol with the RED queue management algorithm. 
The~\ref{sec:Modelica}~section provides general information about the 
physical modeling language Modelica which implements, among other things, 
a hybrid paradigm, and this section also presents the implementation of the 
algorithm in the Modelica language. The~\ref{sec:Julia}~section provides 
information on the Julia programming language and the DifferentialEquations 
package for solving various differential equations. Also in this section some 
features of the Julia language are presented in the context of hybrid modeling by 
using the example of the RED module implementation for active traffic management.

\subsection{Notations and conventions}
\label{sec:notation}

In the work we will use the following variables and notation. The main
 parameters of the system are follows:
\begin{itemize}
\item $w(t)$ is the average TCP window size (in packets);
\item $q(t)$ is the average queue length (in packets);
\item $\hat{q}(t)$ is the exponentially weighted moving average of the queue length;
\item $T(q,t)$ is the Round Trip Time (RTT, sec);
\item $N(t)$~is the number of  TCP sessions;
\item $C$ corresponds to the speed of procession  of packets in the
queue.
\end{itemize}

\section{Random Early Detection as an active queue managment algorithm}
\label{sec:RED}

The authors have repeatedly described the research problems, the
features of the phenomenon under study, the construction of a
mathematical
model~\cite{kulyabov:2019:ecms:red-control,kulyabov:2018:aisc:metodology}.
  
In data transmission networks, there is a global synchronization, 
during which TCP sources send packets synchronously and also 
stop transmission synchronously. Such a stable self-oscillating 
mode of the system operation affects the quality of network service, 
throughput and latency negatively. The use of Active Queue Management 
algorithms, like RED~\cite{floyd:1993:red}, to control traffic reduces the probability of global 
synchronization, but it does not completely eliminate it. For some values 
of the initial parameters in the router settings, self-oscillations of the 
main parameters occur in the system.

An active queue management policy with a RED type algorithm is used 
to control and prevent congestions in the queues of routers. 
Algorithms for managing the state of traffic can be represented 
as control modules in network equipment. The advantage of this 
algorithm is its efficiency and relatively simple implementation 
on network equipment. 

Let us describe a model for transmitting TCP-like traffic in which 
the dynamic flow rate is regulated by an RED type algorithm. 
The model consists of two elements - the packet-generating 
TCP source and the receiver, which is the queue of the router. 
It processes incoming packets in accordance with the control 
algorithm and notifies the source that the packets have reached 
their destination. The source and receiver interact through an 
intermediate link in accordance with the control algorithm, i.e. 
the value of the average queue length affects the source parameters, 
in particular, the size of the TCP congestion window.

A mathematical model of the interaction of an incoming TCP stream 
and a router that processes traffic using a RED-type control algorithm 
is the autonomous system of three differential equations~\cite{misra:2000:fluid-based,misra:1999:sdu,kulyabov:2015:mmg:red,kulyabov:2016:mmcp:one-step}:

\begin{align}
    \dot{w}(t) &=
    \frac{1}{T(t)} \vartheta(w_{\max} - w)
    -
    {\frac{w(t)}{2}\frac{w(t-T(t))}{T(t-T(t))}}p(t-T(t)),
     \label{eq:Window_size}
    \\
    \Dot{q}(t) &=
    \begin{dcases}
      (1-p(t)){\frac{N(t)w(t)}{T(t)}-C}, & q(t)>0,\\
      \max \left[ (1-p(t)){\frac{N(t)w(t)}{T(t)}-C, 0} \right], & q(t)=0,
    \end{dcases}
     \label{eq:Queue}
    \\
    \Dot{\Hat{q}}(t) &= -w_{q} C \Hat{q}(t) + w_{q} C q(t),
     \label{eq:Exp_average}
\end{align}

The~\eqref{eq:Window_size}~equation describes the dynamic change 
of the average TCP window size. The~$1 / T(q,t)$~element reflects 
the slow start TCP state, during which the window size increases per RTT. 
The~$\vartheta (w_{\max} - w)$ component, which is a Heaviside function, 
limits the growth of the TCP window. The~$(1-p(t))w(t)N(t)/T(q,t)$~component 
in the~\eqref{eq:Queue}~equation reflects the increase in the queue 
length when packets arrive which corresponds to the average packet arrival rate.

The~\eqref{eq:Exp_average}~equation reflects the change 
of the $\hat{q}(t)$ - exponentially weighted moving average 
of the queue length function operating as a low-pass filter. It is introduced 
for some smoothing outliers of the queue length. As soon as the~$\hat{q}(t)$~value exceeds 
some predetermined threshold value, the router finds out that the system 
has started overloading and reports this to the source.

The RED algorithm is controlled by the $p(\hat{q})$ - piecewise probabilistic packet drop function 
which can be written as a system of nonlinear equations. 

\begin{equation}
{p}(\hat{q}) =
 \begin{cases}
   0, &\text{ $ 0 \leqslant \hat{q}(t) < q_{\min} $},\\
   \frac{\hat{q}(t) - q_{\min}}{q_{\max} - q_{\min}} p_{\max}  , &\text{ $q_{\min}\leqslant\hat{q}(t)\leqslant q_{\max}$},\\
   1, &\text{ $\hat{q}(t) > q_{\max}$}.
 \end{cases}
 \label{eq:drop_probability}
\end{equation}

  This function depends on the threshold values of the queue size
  $q_{\min}$ and $q_{\max}$, as well as the $p_{\max}$ parameter that
  sets the part of packets that will be discarded if $\hat{q}$ reaches
  the maximum value.

\section{Modeling in the Modelica language}
\label{sec:Modelica}

The Modelica language was developed by the non-profit organization Modelica 
which also develops a freely distributed library based on it. 
This object-oriented language of physical modeling is used to solve 
a wide range of problems~\cite{fritzson:2003,fritzson:2011}. 
Modelica is well suited for component-oriented modeling of complex systems 
consisting of various physical components, also having control components 
and elements oriented to individual processes. 
Let us demonstrate the application of this language 
to hybrid modeling of communication network algorithms~\cite{farnqvist:2002:hybrid_modeling}.

The basis of the language is represented by classes that can be inherited. In Modelica 
classes contain methods and fields that can have types 
of variability such as constant, parameter and variable. In addition to methods and fields, 
the class contains functions and equations which are defined in the~\verb|equation| section. 
One of the mandatory requirements of the Modelica program is the coincident 
number of variables and equations.

Let us declare the variables and parameters used as initial ones.

\begin{minted}{modelica}
Real p(start = 0.0) "Drop probability"; 
Real w(min = 1.0, max = wmax, start = 1.0, fixed = true) "TCP window"; 
Real q(max = R, start = 0.0, fixed = true) "Instant queue length"; 
Real q_avg(start = 0.0) "EWMS queue length";

parameter Real N(start = 60.0) "Number of TCP sessions"; 
parameter Real c(start = 10.0) "Service intensity, Mbps"; 
parameter Real packet_size(start = 500.0) "Package size, bit"; 
parameter Real T(start = 0.05) "RTT"; 
parameter Real thmin(start = 0.25) "Normalized lower threshold"; 
parameter Real thmax(start = 0.5) "Normalized upper threshold"; 
parameter Real R(start = 300.0) "Queue size"; 
parameter Real wq(start = 0.0004) EWMS parameter"; 
parameter Real pmax(start = 0.1) "Maximum drop probability"; 
parameter Real wmax(start = 32.0) "Max window size";
parameter Real C = 125000.0 * c / packet_size "Service intensity, packets"; 
\end{minted}

The congestion control algorithm in the queues of the RED router 
in Modelica is implemented as a \verb|Red| class ~\cite{kulyabov:2019:ceur-ws:2407:red}. 
A discrete drop function is easily set in the main class using the standard \verb|if| operator:

\begin{minted}{modelica}
class Red 
equation 
p = if q_avg < thmin * R 
    then 0.0 
else if q_avg > thmax * R 
    then 1.0 
else (q_avg / R - thmin) * pmax / (thmax - thmin); 
\end{minted}

In this implementation, the discrete element of the system interacts 
well with continuous elements, the behaviour of which is also 
implemented in the main class of the program using three equations:

\begin{minted}{modelica}
equation 
der(w) = wAdd(w, wmax, T) + (-0.5) * w * delay(w, T, T) * delay(p, T, T) / delay(T, T, T); 
der(q) = qAdd(pre(q), w, T, C, N, R); 
der(q_avg) = wq * C * (q - q_avg);
\end{minted}

In Modelica the \verb|der| operator defines a time derivative. 
The delay is realized very simply with the help of the  \verb|delay| operator, 
which makes it possible to work with the delay of both continuous and discrete system elements. 

The \verb|wAdd| function limits the size of the TCP window:

\begin{minted}{modelica}
function wAdd 
input Real wIn, wmax, T; 
output Real wOut; 
algorithm 
wOut := if noEvent(wIn >= wmax) then 0.0 else 1.0 / T; 
end wAdd;
\end{minted}

The \verb|qAdd| function sets the change in the average 
queue length depending on the discrete packet drop function:

\begin{minted}{modelica}
function qAdd 
input Real q, w, T, C, N, R; 
output Real qOut; 
protected Real q1, q2; 
algorithm 
q1 := N * w / T - C; 
q2 := q + q1; 
qOut := if q2 > R then R - q else if q2 > 0.0 then q1 else -q; 
end qAdd;
\end{minted}

Additional restrictions on the size of $w(t)$ and $q(t)$ 
variables are set using the \verb|when| operator in the \verb|Red| class:

\begin{minted}{modelica}
when w <= 1.0
  then reinit(w, 1.0); 
end when; 
when q >= R
  then reinit(q, R); 
end when; 
\end{minted}

\begin{figure}
  \centering
  \includegraphics[width=0.5\linewidth]{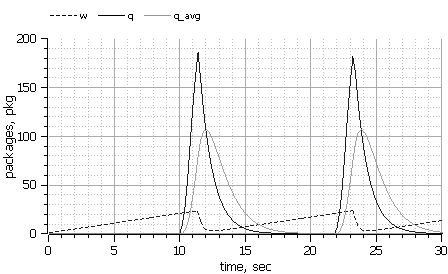}
  \caption{$w(t)$, $q(t)$, $\hat q(t)$ parameters behaviour based on Modelica modeling results}
  \label{fig:Modelica}
\end{figure}

As a result of system modeling the dynamics of the change in 
$w(t)$, $q(t)$, $\hat q(t)$ parameters was obtained, presented 
in figure~\ref{fig:Modelica}. The graph shows that for some 
parameter values in the system there is a stable self-oscillating 
mode of operation.

\section{Modeling in the Julia language}
\label{sec:Julia}

Julia is a high-level language designed for scientific and engineering 
calculations~\cite{farnqvist:2002:hybrid_modeling,bezanson:2012:julia-dinamic-language,bezanson:2017:julia-numeric-computing,joshi:book:learning-julia}.

Let us describe the implementation of an Active Queue Management 
algorithm with a RED control algorithm in Julia\footnote{We used the Julia-1.4.1.}. 

In this implementation, we used the DifferentialEquations
package~\cite{rackauckas:2017:differentialequations.jl}, designed for
effective solution of differential equations of various types, such as
ordinary differential equations, stochastic ordinary differential
equations, differential algebraic and hybrid equations, as well as
delay differential equations.

To install the package we will use the following command in Julia REPL:

\begin{minted}{julia}
using Pkg
Pkg.add("DifferentialEquations")
\end{minted}

We will enable the package using the command:

\begin{minted}{julia}
using DifferentialEquations
\end{minted}

We set the vector of the initial system parameters
\mintinline{julia}{p = (T, N, C, wq, q_min, q_max, R, p_max, w_max)}. The
\mintinline{julia}{pr} variable which is a function of the packet drop probability, 
acts as a global variable.

\begin{minted}{julia}
T = 0.5 
N = 60.0 
C = 10 * 1000000.0 / (8.0 * 500)
wq = 0.0004 
q_min = 0.25 
q_max = 0.50 
R = 300.0 
p_max = 0.1 
w_max = 32.0 
p = (T, N, C, wq, q_min, q_max, 
   R, p_max, w_max)
pr = 0.0
\end{minted}

Since the original system of differential equations contains delayed arguments, 
we define the history function \mintinline{julia}{h(p, t)}
which depends on the parameters vector \mintinline{julia}{p} and time \mintinline{julia}{t}.
Next, we define the delay of the $w(t-T(t))$ variable in \eqref{eq:Window_size}~equation:

\begin{minted}{julia}
h(p, t) = zeros(1)
tau = T
lags = [tau]
\end{minted}

For our task, the \mintinline{julia}{Red} dynamic function describing 
the behaviour of differential equations and setting constraints for $w$, $q$, $\hat q(t)$ parameters in DifferentialEquations 
will have the following form:

\begin{minted}{julia}
function Red(du, u, h, p, t)
w, q, q_avg = u
hist1 = h(p, t - T)[1]
du[1] = 1.0 / T - 
        -(w * hist1 * pr / (2.0 * T))
du[2] = qAdd(q,w,T,C,N,R)
du[3] = -wq * C * q_avg + wq * C * q
end
\end{minted}

In the \mintinline{julia}{Red} function, we also set constraints for the $w$,
$q$, $\hat q(t)$ parameters.

\begin{minted}{julia}
if (w <= 1.0)
    w = 1.0
end
if (q >= R)
    q = R
end
if (q_avg >= R)
    q_avg = R
end
\end{minted}

Next we define the $q$ parameter change function.

\begin{minted}{julia}
function qAdd(q,w,T,C,N,R)
    q1 = N * w / T - C
    q2 = q + q1
    if q2 > R
        return R - q
    elseif q2 > 0
         return q1
    else
        return -q
    end
end
\end{minted}

One of the powerful tools of the DifferentialEquations package is callbacks, 
for which two functions are defined. A condition function is needed to check 
if an event has occurred. The affect function is executed if an event has occurred. 

The discrete packet drop function is implemented as a controller 
which is updated every 0.01 seconds until the \mintinline{julia}{tf} simulation time is reached. 
The \mintinline{julia}{tstops} array defines the sampling intervals of the condition check, 
the function checks if \mintinline{julia}{t} is one of the sample element.

\begin{minted}{julia}
tf = 30.0
tstops = collect(0:0.01:tf)
function condition_control_loop(u,t,integrator)
    (t in tstops)
end
\end{minted}

Next, we determine the affect function, which is the controller. 
The \mintinline{julia}{control_loop!} function at each step calculates 
the new value of the packet drop probability  function $p$ depending on the current values of the $w$, $q$, $\hat q(t)$ 
parameters in accordance with the formula~\eqref{eq:drop_probability}. 

\begin{minted}{julia}
function control_loop!(integrator)
    global pr

    w = integrator.u[1]
    q = integrator.u[2]
    q_avg = integrator.u[3]

    if (q_avg < q_min * R)
        pr = 0.0
    elseif (q_avg > q_max * R)
        pr = 1.0
    else pr = p_max * (q_avg/R - q_min) / (q_max - q_min)
    end
end
\end{minted}

Then we define a discrete type callback:

\begin{minted}{julia}
cb = DiscreteCallback(condition_control_loop, control_loop!)
\end{minted}

Finally, we define the vector of the system initial state, the simulation time 
and call the DDEProblem package solver. The \mintinline{julia}{Red} function, vector of initial states, 
simulation time, and variables delay parameters are passed to its arguments:

\begin{minted}{julia}
u0 = [1.0, 0.0, 0.0]
tspan = (0.0, tf)
prob = DDEProblem(Red, u0, h, tspan, p, constant_lags=lags)
alg = MethodOfSteps(Tsit5())
sol = solve(prob, alg, callback = cb, tstops=tstops)
\end{minted}

\begin{figure}
  \centering
  \includegraphics[width=0.5\linewidth]{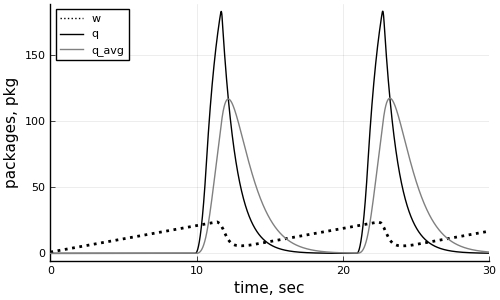}
  \caption{$w(t)$, $q(t)$, $\hat q(t)$ parameters behaviour based on Julia modeling results}
  \label{fig:Atom}
\end{figure}

As a result of the simulation, we obtain a graph demonstrating the behaviour of 
the TCP Reno window size and the average queue length, i.e. reflecting
the dynamics of a queue in a router (or gateway) with a queue
management module using the RED algorithm (fig.~\ref{fig:Atom}).

\section{Results}
\label{sec:results}

As a result of the study, two software complexes 
created in the programming languages Modelica and Julia were obtained.
Both of these complexes implement the same mathematical model of a data 
transmission network with an active traffic control module operating according 
to the RED algorithm.

A numerical experiment conducted within the framework 
of both software systems gives comparable results(see fig.~\ref{fig:Modelica} and
fig.~\ref{fig:Atom}).

\section{Discussion}
\label{sec:discussion}

Modelica and Julia languages are both domain-specific languages. 
However, Julia is considered as the common language of scientific calculations, 
Modelica is considered as a specialized language for modeling of dynamic, 
continuous and hybrid systems. 

The mathematical model of the RED algorithm is formulated 
using a hybrid approach. Therefore, in the framework of the Modelica language, 
its implementation turned out to be quite simple. In this language 
ordinary differential equations are written with a delayed argument 
with the help of high-level tools quite easily. In addition, the use of 
discrete elements is implemented very clearly.

Julia language is aimed at solving a wider range of tasks. 
Therefore, it does not have as many syntax tools as the more specialized Modelica. 
In particular, the implementation of the hybrid paradigm in Julia requires 
a higher qualification of the programmer than when working with the Modelica language. 

For specialized applications, Julia requires a greater level 
of knowledge than specialized languages such as Modelica. Julia is also a metalanguage and it 
can serve as the basis for the construction of other languages. 
For example, the Modia extension~\cite{elmqvist:2016:modia,otter:2017:modia} was made to migrate programs 
from Modelica to Julia (and possibly in the opposite direction).

\section{Conclusion}
\label{sec:conclusion}

The authors demonstrated the application of 
the continuous-discrete approach to the modeling of non-linear systems with control. 
The simulated system is the system consisting of an incoming stream processed 
according to the TCP protocol, as well as a router that processes 
traffic using a RED algorithm.

By comparing software implementations in the Modelica and Julia programming languages, 
the simplicity of  hybrid systems modeling in Modelica is demonstrated, 
where continuous system elements, implemented by using a system of differential 
equations, interact with a discrete packet drop function quite well. 
Restrictions for some system parameters are set using the \verb|when| operator. 
Implementation of the delay is also possible for system 
elements of any nature of functioning.

Julia also provides the ability to model systems 
according to a hybrid paradigm. The DifferentialEquations package 
allows to solve systems of differential equations of various kinds, 
including delay differential equations. The continuous probabilistic 
function has been successfully implemented using the callback option. 
A delayed argument to a continuous function is implemented using the \verb|h(p, t)| history function.

Thus, the authors studied the capabilities of the Modelica and Julia programming 
languages in modeling of hybrid systems containing both continuous and discrete 
aspects of behaviour. The numerical simulation of the process of transmitting 
data via TCP and the process of regulating the flow state using the RED 
algorithm in the event of congestion has been performed, graphs 
demonstrating changes in the main parameters of the system have 
been obtained.

\begin{acknowledgments}

The publication has been
funded by Russian Foundation for Basic Research (RFBR) according to
the research project No~19-01-00645.

\end{acknowledgments}

 \bibliographystyle{elsarticle-num}

\bibliography{bib/modelica-vs-julia/cite}

\end{document}


\graphicspath{{image/modelica-vs-julia/ru/}{image/modelica-vs-julia/}{image/}}

\title{Возможности гибридного моделирования нелинейных систем с управлением в языках Modelica и Julia}

\author{А. М. Ю. Апреутесей}
\email{1032193049@pfur.ru}
\affiliation{Кафедра прикладной информатики и теории вероятностей,\\
  Российский университет дружбы народов,\\
  117198, Москва, ул. Миклухо-Маклая, д. 6}

\author{А. В. Королькова}
\email{korolkova-av@rudn.ru}
\affiliation{Кафедра прикладной информатики и теории вероятностей,\\
  Российский университет дружбы народов,\\
  117198, Москва, ул. Миклухо-Маклая, д. 6}

\author{Д. С. Кулябов}
\email{kulyabov-ds@rudn.ru}
\affiliation{Кафедра прикладной информатики и теории вероятностей,\\
  Российский университет дружбы народов,\\
  117198, Москва, ул. Миклухо-Маклая, д. 6}
\affiliation{Лаборатория информационных технологий,\\
  Объединённый институт ядерных исследований,\\
  ул. Жолио-Кюри 6, Дубна, Московская область, Россия, 141980}

\begin{abstract}

Данная работа посвящена изучению возможностей языков программирования Modelica и Julia для реализации непрерывно-дискретной парадигмы при моделировании гибридных систем, содержащих как непрерывные, так и дискретные аспекты поведения. В качестве моделируемой пороговой системы выступает система, состоящая из входящего потока, обрабатываемого согласно протоколу Transmission Control Protocol (TCP), а также маршрутизатора, обрабатывающего трафик по алгоритму типа Random Early Detection (RED). 

\end{abstract}

  \keywords{активное управление трафиком, имитационное моделирование,
    Modelica, Julia,
    Random Early Detection}

\maketitle

\section{Введение}
\label{sec:intro}

Среди возможных методов исследования сложных систем можно выделить
построение дискретно--событийной модели, построение непрерывной
модели, а также гибридное моделирование~\cite{kulyabov:2016:ecms:red}. В подобных гибридных системах
сочетается работа непрерывных и дискретных элементов, например,
системы с дискретным устройством управления и объектом управления с
непрерывным характером функционирования~\cite{farnqvist:2002:hybrid_modeling}. В качестве исследуемой
системы выступает модель взаимодействия процесса передачи данных по
протоколу Transmission Control Protocol (TCP) и процесса регулирования
состояния потока при возникновении перегрузок, в качестве которого
рассматривается алгоритм Random Early Detection (RED). При
моделировании TCP-подобного трафика можно воспользоваться жидкостным
(непрерывным) подходом, однако особенно важно учитывать дискретные
переходы между TCP состояниями и функцию сброса пакетов в алгоритмах
типа RED, гибридный подход отразит эти важные особенности моделируемой
системы.

  \subsection{Структура статьи}
\label{sec:structure}

Структура статьи следующая. В разделе~\ref{sec:RED} описываются
основные принципы и математическая модель функционирования механизма
управления перегрузками в протоколе TCP с алгоритмом управления
очередями RED. В разделе~\ref{sec:Modelica} дается общая информация об
языке физического моделирования Modelica, реализующем в том числе
гибридную парадигму, представлена реализация алгоритма на языке
Modelica. Раздел~\ref{sec:Julia} дает информацию о языке
программирования Julia и о библиотеке DifferentialEquations для
решения дифференциальных уравнений различных видов. Также в данном
разделе представлены некоторые возможности языка Julia в контексте
гибридного моделирования на примере реализации модуля активного
управления трафиком RED.

\subsection{Обозначения и соглашения}
\label{sec:notation}

В работе будем использовать следующие переменные и
обозначения. Основные параметры рассматриваемой системы:
\begin{itemize}
\item $w(t)$~--- средний размера TCP-окна (в пакетах);
\item $q(t)$~--- средний размер очереди (в пакетах);
\item $\hat{q}(t)$~--- экспоненциально взвешенное скользящее среднее значения длины очереди;
\item $T(q,t)$~--- время двойного оборота (Round Trip Time, RTT, сек);
\item  $N(t)$~--- число TCP-сессий;
\item  $C$~--- скорость обработки пакетов в очереди.
\end{itemize}

\section{Алгоритм активного управления очередью Random Early Detection}
\label{sec:RED}

Авторами неоднократно описывались проблематика исследования,
особенности изучаемого явления, построение математической
модели~\cite{kulyabov:2019:ecms:red-control,kulyabov:2018:aisc:metodology}.
  
В сетях передачи данных имеет место глобальная синхронизация, при
которой TCP источники синхронно отправляют пакеты и также синхронно
прекращают передачу. Такой устойчивый автоколебательный режим работы
системы отрицательным образом сказывается на показателях качества
обслуживания сети, на пропускной способности и
латентности. Использование для управления трафиком алгоритмов
активного управления очередью, таких как RED, снижает вероятность
возникновения глобальной синхронизации, но не устраняет её полностью,
при некоторых значениях начальных параметров в настройках
маршрутизаторов в системе возникают автоколебания основных
параметров.

Алгоритм активного управления очередью с алгоритмом управления типа
RED используется для контроля и предотвращения перегрузок в очередях
маршрутизаторов~\cite{floyd:1993:red}. Алгоритмы управления состоянием
трафика могут быть представлены как модули управления в сетевом
оборудовании. Преимуществом данного алгоритма является его
эффективность и относительно простая реализация на сетевом
оборудовании.

Опишем модель передачи TCP-подобного трафика с регулируемой алгоритмом
типа RED динамической интенсивностью потока. Модель состоит из двух
элементов~--- генерирующего пакеты TCP-источника и получателя, в
качестве которого выступает очередь маршрутизатора, обрабатывающая
поступившие пакеты в соответствии с алгоритмом управления и
уведомляющая источник о доставке пакетов. Источник и получатель
взаимодействуют через промежуточное звено в соответствии с алгоритмом
управления, т.е. значение средней длины очереди влияет на параметры
источника, в частности, на размер окна перегрузки TCP.

Математическая модель взаимодействия входящего TCP-потока и
маршрутизатора, обрабатывающего трафик по алгоритму управления типа
RED, представляет собой автономную систему трёх дифференциальных
уравнений~\cite{misra:2000:fluid-based,misra:1999:sdu,kulyabov:2015:mmg:red,kulyabov:2016:mmcp:one-step}:

\begin{align}
    \dot{w}(t) &=
    \frac{1}{T(t)} \vartheta(w_{\max} - w)
    -
    {\frac{w(t)}{2}\frac{w(t-T(t))}{T(t-T(t))}}p(t-T(t)),
     \label{eq:Window_size}
    \\
    \Dot{q}(t) &=
    \begin{dcases}
      (1-p(t)){\frac{N(t)w(t)}{T(t)}-C}, & q(t)>0,\\
      \max \left[ (1-p(t)){\frac{N(t)w(t)}{T(t)}-C, 0} \right], & q(t)=0,
    \end{dcases}
     \label{eq:Queue}
    \\
    \Dot{\Hat{q}}(t) &= -w_{q} C \Hat{q}(t) + w_{q} C q(t),
     \label{eq:Exp_average}
\end{align}

Уравнение \eqref{eq:Window_size} описывает динамическое изменение
среднего размера TCP окна $w(t)$. Элемент $1 / T(q,t)$ отражает фазу
медленного старта TCP, во время которой размер окна увеличивается за единицу времени двойного оборота.
Компонента $\vartheta (w_{\max} - w)$, являющаяся функцией
Хэвисайда, ограничивает роста TCP-окна.

Компонента $(1-p(t))w(t)N(t)/T(q,t)$ в уравнении \eqref{eq:Queue}
отражает увеличение длины очереди при поступлении пакетов, что
соответствует средней интенсивности поступления пакетов.

Уравнение \eqref{eq:Exp_average} отражает изменение функции
экспоненциально взвешенного скользящего среднего значения длины
очереди $\hat{q}(t)$, которое вводится для некоторого сглаживания
выбросов мгновенного значения размера очереди $q(t)$, функционируя как
фильтр низких частот. Как только значение $\hat{q}(t)$ превышает некое
заданное пороговое значение, маршрутизатор узнает о том, что в системе
началась перегрузка и сообщает об этом источнику.

Непосредственно за управление по алгоритму RED отвечает кусочная
вероятностная функция сброса пакета $p(\hat{q})$, которая может быть
записана как система нелинейных уравнений.

\begin{equation}
{p}(\hat{q}) =
 \begin{cases}
   0, &\text{ $ 0 \leqslant \hat{q}(t) < q_{\min} $},\\
   \frac{\hat{q}(t) - q_{\min}}{q_{\max} - q_{\min}} p_{\max}  , &\text{ $q_{\min}\leqslant\hat{q}(t)\leqslant q_{\max}$},\\
   1, &\text{ $\hat{q}(t) > q_{\max}$}.
 \end{cases}
 \label{eq:drop_probability}
\end{equation}

Данная функция зависит от пороговых значений размера
очереди $q_{\min}$ и $q_{\max}$, а также параметра $p_{\max}$,
задающего часть пакетов, которые будут отброшены в случае, если
$\hat{q}$ достигнет максимального значения.

\section{Моделирование на языке Modelica}
\label{sec:Modelica}

Язык Modelica разработан некоммерческой организацией Modelica, которая
также разрабатывает на его основе свободно распространяемую библиотеку. Этот язык,
позиционируемый как объектно-ориентированный язык физического
моделирования, применяется для решения широкого круга задач~\cite{fritzson:2003,fritzson:2011}. Modelica
хорошо применима для компонентно-ориентированного моделирования
сложных систем, состоящих из различных физических компонентов, также
имеющих компоненты управления и элементы, ориентированные на отдельные
процессы. Продемонстрируем применение данного языка к гибридному
моделированию алгоритмов сетей связи~\cite{farnqvist:2002:hybrid_modeling}.

Основу языка составляют имеющие возможность наследоваться классы,
которые содержат в себе все элементы наследуемого класса. В Modelica
классы содержат методы и поля, которые могут иметь такие типы
изменчивости как константа, параметр и переменная. Помимо методов и
полей в классе содержатся функции и связывающие переменные уравнения,
которые задаются в разделе \verb|equation|. Одним из обязательных требований
программы на Modelica является совпадающее число переменных и
уравнений.

Объявим переменные и приведем параметры, используемые в качестве начальных. 

\begin{minted}{modelica}
Real p(start = 0.0) "Вероятность сброса"; 
Real w(min = 1.0, max = wmax, start = 1.0, fixed = true) "Окно"; 
Real q(max = R, start = 0.0, fixed = true) "Мгновенная длина очереди"; 
Real q_avg(start = 0.0) "EWMS длины очереди";

parameter Real N(start = 60.0) "Количество сессий"; 
parameter Real c(start = 10.0) "Интенсивность обслуживания, Mbps"; 
parameter Real packet_size(start = 500.0) "Размер пакета, bit"; 
parameter Real T(start = 0.05) "Время двойного оборота"; 
parameter Real thmin(start = 0.25) "Нормализованный нижний порог"; 
parameter Real thmax(start = 0.5) "Нормализованный верхний порог"; 
parameter Real R(start = 300.0) "Размер очереди"; 
parameter Real wq(start = 0.0004) "Параметр EWMS"; 
parameter Real pmax(start = 0.1) "Максимальная вероятность сброса"; 
parameter Real wmax(start = 32.0) "Максимальный размер окна";
parameter Real C = 125000.0 * c / packet_size "Интенсивность обслуживания, packets"; 
\end{minted}

Алгоритм контроля перегрузки в очередях маршрутизатора RED на языке
Modelica реализован в виде класса
\verb|Red|~\cite{kulyabov:2019:ceur-ws:2407:red}. Дискретная функция
сброса легко задается в основном классе с помощью стандартного
оператора \verb|if|:

\begin{minted}{modelica}
class Red 
equation 
p = if q_avg < thmin * R 
    then 0.0 
else if q_avg > thmax * R 
    then 1.0 
else (q_avg / R - thmin) * pmax / (thmax - thmin); 
\end{minted}

В данной реализации алгоритма RED дискретный элемент системы хорошо
взаимодействует с непрерывными элементами, поведение которых также
реализовано в основном классе программы с помощью трех уравнений:

\begin{minted}{modelica}
equation 
der(w) = wAdd(w, wmax, T) + (-0.5) * w * delay(w, T, T) * delay(p, T, T) / delay(T, T, T); 
der(q) = qAdd(pre(q), w, T, C, N, R); 
der(q_avg) = wq * C * (q - q_avg);
\end{minted}

На языке Modelica оператор \verb|der| задает производную по
времени. Запаздывание в Modelica реализуется крайне просто с помощью
оператора \verb|delay|, который дает возможность работать с
запаздыванием как непрерывных, так и дискретных элементов системы.

Функция \verb|wAdd| ограничивает рост размера TCP-окна:

\begin{minted}{modelica}
function wAdd 
input Real wIn, wmax, T; 
output Real wOut; 
algorithm 
wOut := if noEvent(wIn >= wmax) then 0.0 else 1.0 / T; 
end wAdd;
\end{minted}

Функция \verb|qAdd| задает изменение среднего размера очереди в
зависимости от дискретной функции сброса пакетов:

\begin{minted}{modelica}
function qAdd 
input Real q, w, T, C, N, R; 
output Real qOut; 
protected Real q1, q2; 
algorithm 
q1 := N * w / T - C; 
q2 := q + q1; 
qOut := if q2 > R then R - q else if q2 > 0.0 then q1 else -q; 
end qAdd;
\end{minted}

Дополнительные ограничения в размере переменных $w(t)$ и $q(t)$
задаются с помощью оператора \verb|when| в классе \verb|Red|:

\begin{minted}{modelica}
when w <= 1.0
  then reinit(w, 1.0); 
end when; 
when q >= R
  then reinit(q, R); 
end when; 
\end{minted}

\begin{figure}
  \centering
  \includegraphics[width=0.5\linewidth]{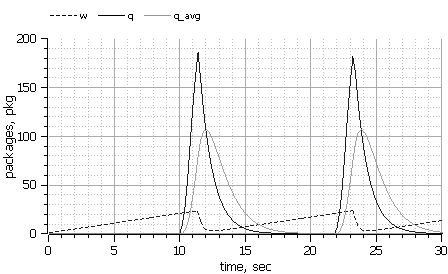}
  \caption{Поведение параметров $w(t)$, $q(t)$, $\hat q(t)$ по результатам моделирования на языке Modelica}
  \label{fig:Modelica}
\end{figure}

В результате моделирования системы была получена динамика изменения
$w(t)$, $q(t)$, $\hat q(t)$, представленная на
рис.~\ref{fig:Modelica}. График демонстрирует, что при некоторых
значениях параметров в системе возникает устойчивый автоколебательный
режим функционирования.

\section{Моделирование на языке Julia}
\label{sec:Julia}

Язык Julia --- это язык высокого уровня, предназначенный для научных и
инженерных
расчётов~\cite{farnqvist:2002:hybrid_modeling,bezanson:2012:julia-dinamic-language,bezanson:2017:julia-numeric-computing,joshi:book:learning-julia}.

Опишем реализацию алгоритма активного управления очередью с алгоритмом
управления типа RED на языке Julia\footnote{Мы использовали версию Julia 1.4.1.}.

В данной реализации использовалась библиотека DifferentialEquations~\cite{rackauckas:2017:differentialequations.jl},
предназначенная для эффективного решения дифференциальных уравнений
различных видов, таких как обыкновенные дифференциальные уравнения,
стохастические обыкновенные дифференциальные уравнения,
дифференциально-алгебраические и гибридные уравнения, а также
дифференциальные уравнения с запаздыванием.

Для установки пакета используем следующую команду в Julia REPL:

\begin{minted}{julia}
using Pkg
Pkg.add("DifferentialEquations")
\end{minted}

Подключим пакет, используя команду:

\begin{minted}{julia}
using DifferentialEquations
\end{minted}

Зададим вектор начальных параметров системы
\mintinline{julia}{p = (T, N, C, wq, q_min, q_max, R, p_max, w_max)}. Переменная
\mintinline{julia}{pr}, являющаяся функцией вероятности сброса пакетов, выступает
как глобальная переменная.

\begin{minted}{julia}
T = 0.5 
N = 60.0 
C = 10 * 1000000.0 / (8.0 * 500)
wq = 0.0004 
q_min = 0.25 
q_max = 0.50 
R = 300.0 
p_max = 0.1 
w_max = 32.0 
p = (T, N, C, wq, q_min, q_max, 
   R, p_max, w_max)
pr = 0.0
\end{minted}

Так как в исходной системе дифференциальных уравнений присутствуют
запаздывающие аргументы, определим функцию истории \mintinline{julia}{h(p, t)},
которая зависит от вектора параметров \mintinline{julia}{p} и времени \mintinline{julia}{t}.
Далее определим запаздывание переменной $w(t-T(t))$ в
уравнении~\eqref{eq:Window_size}:

\begin{minted}{julia}
h(p, t) = zeros(1)
tau = T
lags = [tau]
\end{minted}

Для предложенной нами задачи динамическая функция \mintinline{julia}{Red},
описывающая поведение дифференциальных уравнений и задающая
ограничения для параметров $w$, $q$, $\hat q(t)$, в
DifferentialEquations будет иметь следующий вид:

\begin{minted}{julia}
function Red(du, u, h, p, t)
w, q, q_avg = u
hist1 = h(p, t - T)[1]
du[1] = 1.0 / T - 
        -(w * hist1 * pr / (2.0 * T))
du[2] = qAdd(q,w,T,C,N,R)
du[3] = -wq * C * q_avg + wq * C * q
end
\end{minted}

В функции \mintinline{julia}{Red} также зададим ограничения для параметров $w$,
$q$, $\hat q(t)$.

\begin{minted}{julia}
if (w <= 1.0)
    w = 1.0
end
if (q >= R)
    q = R
end
if (q_avg >= R)
    q_avg = R
end
\end{minted}

Далее зададим функцию изменения параметра $q$.

\begin{minted}{julia}
function qAdd(q,w,T,C,N,R)
    q1 = N * w / T - C
    q2 = q + q1
    if q2 > R
        return R - q
    elseif q2 > 0
         return q1
    else
        return -q
    end
end
\end{minted}

Одним из мощных инструментов пакета DifferentialEquations являются
обратные вызовы (callbacks), для работы с которыми определяются две
функции. Функция условия (condition function) необходима для проверки
того, произошло ли некоторое событие. Воздействующая функция (affect
function) будет выполняться, если событие
произошло.

Дискретная функция сброса пакетов реализуется как контроллер, который
обновляется каждые 0.01 сек до достижения времени моделирования
\mintinline{julia}{tf}. Массив \mintinline{julia}{tstops} определяет интервалы выборки проверки
условия, функция \mintinline{julia}{condition_control_loop} проверяет является ли
\mintinline{julia}{t} одним из экземпляров выборки:

\begin{minted}{julia}
tf = 30.0
tstops = collect(0:0.01:tf)
function condition_control_loop(u,t,integrator)
    (t in tstops)
end
\end{minted}

Далее определим функцию воздействия, которая и является
контроллером. Функция \mintinline{julia}{control_loop!} на каждом шаге вычисляет
новое значение вероятностной функции сброса пакетов $p$ в зависимости
от текущих значений параметров $w$, $q$, $\hat q(t)$ в соответствии с
формулой~\eqref{eq:drop_probability}:

\begin{minted}{julia}
function control_loop!(integrator)
    global pr

    w = integrator.u[1]
    q = integrator.u[2]
    q_avg = integrator.u[3]

    if (q_avg < q_min * R)
        pr = 0.0
    elseif (q_avg > q_max * R)
        pr = 1.0
    else pr = p_max * (q_avg/R - q_min) / (q_max - q_min)
    end
end
\end{minted}

Зададим обратный вызов дискретного типа:

\begin{minted}{julia}
cb = DiscreteCallback(condition_control_loop, control_loop!)
\end{minted}

Наконец, определим вектор начального состояния системы, время
моделирования и вызовем решатель пакета DDEProblem, в аргументы
которого передается функция \mintinline{julia}{Red}, вектор начальных состояний
системы, время моделирования и параметры задержки переменных:

\begin{minted}{julia}
u0 = [1.0, 0.0, 0.0]
tspan = (0.0, tf)
prob = DDEProblem(Red, u0, h, tspan, p, constant_lags=lags)
alg = MethodOfSteps(Tsit5())
sol = solve(prob, alg, callback = cb, tstops=tstops)
\end{minted}

\begin{figure}
  \centering
  \includegraphics[width=0.5\linewidth]{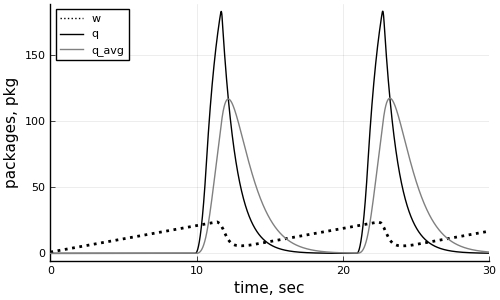}
  \caption{Поведение параметров $w(t)$, $q(t)$, $\hat q(t)$ по результатам моделирования на языке Julia}
  \label{fig:Atom}
\end{figure}

В результате моделирования получим график изменения размера окна TCP
Reno, отражающий динамику управления перегрузкой TCP, а также график среднего
размера очереди, отражающий динамику очереди в маршрутизаторе (или
шлюзе) с модулем управления очередью по алгоритму RED (рис.~\ref{fig:Atom}).

\section{Результаты}
\label{sec:results}

В результате исследования получены два программных комплекса,
созданных на языках программирования Modelica и Julia. Оба этих
комплекса реализуют одну и ту же математическую модель сети передачи
данных с модулем активного управления трафиком, работающим по
алгоритму RED.

Численный эксперимент, проведённый в рамках обоих программных
комплексов, даёт сопоставимые результаты (см. рис.~\ref{fig:Modelica} и
рис.~\ref{fig:Atom}).

\section{Обсуждение}
\label{sec:discussion}

Оба языка программирования, и Modelica, и Julia являются
предметно-ориентированными языками. Впрочем, Julia при этом
рассматривается как общий язык научных расчётов, а Modelica как
специализированный язык моделирования динамических систем, непрерывных
и гибридных.

Математическая модель алгоритма RED сформулирована в рамках гибридного
подхода. Поэтому в рамках языка Modelica её реализация вышла
достаточно простой. На этом языке очень естественно, с помощью
высокоуровневых средств производится запись обыкновенных дифференциальных уравнений с
запаздывающим аргументом. Кроме того, использование дискретных
элементов реализовано весьма наглядно.

Язык Julia направлен на решение более широкого круга задач. Поэтому
он не обладает таким количеством синтаксического сахара, как более
специализированная Modelica. В частности, реализация гибридной
парадигмы в Julia требует более высокой квалификации программиста, чем
при работе с языком Modelica. 

Для специализированного применения Julia требует большего уровня
знаний, нежели специализированные языки, такие, как Modelica. Впрочем,
Julia является также и метаязыком, и может служить основой для
конструирования других языков. Например, расширение
Modia~\cite{elmqvist:2016:modia,otter:2017:modia} было сделано для
миграции программ с Modelica на Julia (и, возможно, и в обратном
направлении).

\section{Заключение}
\label{sec:conclusion}

Авторами было продемонстрировано применение непрерывно-дискретного
подхода к моделированию нелинейных систем с управлением. В качестве
моделируемой системы выступала система, состоящая из входящего
потока, обрабатываемого согласно протоколу TCP, а также
маршрутизатора, обрабатывающего трафик по алгоритму типа RED.

В результате сравнения программных реализаций на языках
программирования Modelica и Julia продемонстрирована простота
моделирования гибридных систем в Modelica, где непрерывные элементы
системы, реализованные с помощью системы дифференциальных уравнений,
хорошо взаимодействует с дискретной функцией сброса
пакетов. Ограничения для некоторых параметров системы задается с
помощью оператора \verb|when|. Реализация запаздывания также возможна
для элементов системы любого характера функционирования.

Julia также дает возможность моделировать системы согласно гибридной
парадигме. Пакет DifferentialEquations позволяет решать системы
дифференциальных уравнений различных видов, в том числе и
дифференциальные уравнения с запаздыванием. Непрерывная вероятностная
функция успешно реализована с использованием опции обратных
вызовов. Запаздывающий аргумент непрерывной функции был реализован с
помощью функции истории \verb|h(p, t)|. 

Таким образом, авторами были изучены возможности языков
программирования Modelica и Julia при моделировании гибридных систем,
содержащих как непрерывные, так и дискретные аспекты
поведения. Проведено численное моделирование процесса передачи данных
по протоколу TCP и процесса регулирования  алгоритмом RED состояния потока при
возникновении перегрузок, получены графики,
демонстрирующие изменения основных параметров системы.

\def\acknowledgmentsname{Благодарности}

\begin{acknowledgments}

Публикация подготовлена при
при финансовой поддержке РФФИ в рамках научного проекта
\textnumero~19-01-00645.

\end{acknowledgments}

  \ifthenelse{\boolean{xetex} \OR \boolean{luatex}}
  {\bibliographystyle{ugost2008l}}
  {\bibliographystyle{gost2008l}}

\bibliography{bib/modelica-vs-julia/cite}